# Simulation of Polyelectrolytes in Solution Using Dissipative Particle Dynamics in the Grand Canonical Ensemble: Interaction Strength and Salt Effects.

F. Alarcón Oseguera [1][2], E. Pérez[1] and A. Gama Goicochea[2]*

*[1] Instituto de Física, Universidad Autónoma de San Luis Potosí, Álvaro Obregón 64, 78000 San Luis Potosí, México.*

*[2] Centro de Investigación en Polímeros, COMEX Group, Marcos Achar Lobatón 2, 55885 Tepexpan, Edo. De México, México.*

We have studied a bulk electrolyte, and polyelectrolyte solutions with surfactants or multivalent salt with the explicit presence of counterions and solvent molecules by means of the mesoscopic dissipative particle dynamics (DPD) method in the Grand Canonical ensemble. The electrostatic interactions are calculated using the Ewald sum method and the structure of the fluid is analyzed through the radial distribution function between charged particles. The results are in very good agreement with those reported in the literature using a different method for the calculation of the electrostatic forces, and with those obtained using DPD in the canonical ensemble. We also studied the salt dependent conformation of polyelectrolyte solutions as a function of the solvent quality, and analyzed the electrostatic interaction strength dependence of dilute flexible polyelectrolytes in solution. For the complex systems mentioned above, the electrostatic interactions and the solvent quality play a key role in understanding phenomena that do not occur in noncharged systems.

## INTRODUCTION

Charged polymers are macromolecules with ionizable groups. That means that these groups can dissociate, leaving charges on polymer chains and releasing counterions into a polar solvent such as water. Charged polymers that carry only acidic or basic groups are known as polyelectrolytes[1]. They have practical importance for their applications as processing aids such as flocculants, dewatering agents, demulsifiers, as drag reduction agents; as additives in detergents and cosmetics; in the manufacture of membranes, ion-exchange resins, gels, and modified plastics. They are also important as biological molecules such as proteins, sugars, and nucleic acids. Therefore polyelectrolytes are an active and extensively reviewed research field[2].

Atom-based simulations such as the molecular dynamics (MD) method have come into wide use for material design; however MD has limitations in the time and length scales involved in the simulation, as it becomes prohibitively expensive for large systems. As an alternative to atomistic simulations there exist mesoscopic simulation methods like DPD that was developed to reach larger lengths and longer time scales than MD. This kind of mesoscopic approach has been applied to study a wide variety of complex fluids including polymers. However, the prediction of the conformational properties of a charged polymer is a very complex problem due to the long range nature of the Coulomb interaction. A well known technique to solve the problem of the long range interactions in computer simulations is by means of the Ewald sums method.

* Corresponding author. Email: agama@cip.org

In this paper we present Monte Carlo (MC) simulations in the Grand Canonical (GC) ensemble. It should be pointed out that we could use any other ensemble such as the Canonical one, since in the thermodynamic limit different ensembles for the same system should lead to the same physical results[3]. Although computer simulations are never truly performed in the thermodynamic limit, for relatively short ranged forces the use of a particular ensemble becomes a matter of physical or mathematical convenience for the particular problem under consideration. We therefore choose to work in the GC ensemble, having in mind future applications such as confined systems in equilibrium with a bulk, where the chemical potential must be fixed to ascertain that chemical, as well as thermodynamic equilibrium is achieved, which is where the GC ensemble is the best choice[4].

The MC in the GC ensemble we developed is hybridized with DPD and takes into account the charge of the particles explicitly, for the prediction of conformational properties of polyelectrolytes. We describe in Section I the simulation method as well as MC in the GC ensemble (GCMC). The Ewald method and the charge distribution function that we used are described in Section II. Then, to validate our methodology, we present in Section III the radial distribution functions of a bulk electrolyte and a polyelectrolyte – surfactant solution compared with results obtained with other techniques. This methodology is then used to analyze the size of a polyelectrolyte chain as a function of the Bjerrum length, taking into account the electrostatic interaction between monomers and the counterions immerse in a solution without salt. The results are presented in Section IV. Additionally, in Section V we present and discuss results on the size of a polyelectrolyte, using the concept of radius of gyration as a function of multivalent salt concentration. The scaling exponent as a function of the salt concentration in theta solvent is also presented. As shown there, the solvent quality plays an important role in the contraction and reexpansion of the polyelectrolytes. Finally we present some conclusions and perspectives of the work in Section VI.

## I. SIMULATION METHOD

The DPD method has been used successfully in several applications related to systems with mesoscopic length scales[4-6]. Since its introduction[7] many theoretical advances and modifications have been made to put the structure of DPD on a firm foundation[8-10] and have even combined this coarse grained dynamics with some other techniques[11,12].

We have applied the DPD technique hybridized with GCMC to study the properties of charged polymer solutions, where the DPD beads represent a mesoscopic region of fluid. The DPD equations of motion of each particle are subjected to the following stochastic differential equations:

$$\dot{\boldsymbol{v}}_i = \sum_{j \neq i} \left( \boldsymbol{F}_{ij}^C + \boldsymbol{F}_{ij}^D + \boldsymbol{F}_{ij}^R \right) \tag{1}$$

$$\dot{\boldsymbol{r}}_i = \boldsymbol{v}_i, \tag{2}$$

for simplicity all masses are taken as equal to 1. The forces in equation (1) can be described as conservative $\boldsymbol{F}_{ij}^C$, dissipative $\boldsymbol{F}_{ij}^D$ and random $\boldsymbol{F}_{ij}^R$, and have the forms

$$\boldsymbol{F}_{ij}^C = a_{ij} \omega^C(r_{ij}) \hat{\boldsymbol{e}}_{ij} \tag{3}$$

$$\boldsymbol{F}_{ij}^D = -\gamma \omega^D(r_{ij}) [\hat{\boldsymbol{e}}_{ij} \cdot \boldsymbol{v}_{ij}] \hat{\boldsymbol{e}}_{ij} \tag{4}$$

$$\boldsymbol{F}_{ij}^R = \sigma \omega^R(r_{ij}) \hat{\boldsymbol{e}}_{ij} \xi_{ij} \tag{5}$$

where $\boldsymbol{r}_{ij} = \boldsymbol{r}_i - \boldsymbol{r}_j$ is the relative separation vector, $\hat{\boldsymbol{e}}_{ij}$ is the unit vector in the direction of $\boldsymbol{r}_{ij}$, and $\boldsymbol{v}_{ij} = \boldsymbol{v}_i - \boldsymbol{v}_j$ where $\boldsymbol{v}_i$ is the velocity of particle *i*; $a_{ij}$ is the maximum repulsion between particles *i* and *j*. The random variable $\xi_{ij}$ is Gaussian white noise with the stochastic properties of zero mean: $\langle \xi_{ij} \rangle = 0$ and the noncorrelation for different pairs of particles and different times: $\langle \xi_{ij}(t) \xi_{i'j'}(t') \rangle = \left( \delta_{ii'} \delta_{jj'} + \delta_{ij'} \delta_{ji'} \right) \delta(t - t')$. With these assumptions, Español and Warren[8] demonstrated that, in order to obtain a canonical distribution function, the constants in equations (4) and (5) must satisfy:

$$\sigma = (2 k_B T \gamma)^{1/2} \tag{6}$$

with $k_B$ equal to Boltzmann's constant, *T* is the absolute temperature and the weight functions only need to fulfill that

$$\omega^D(r_{ij}) = \left[ \omega^R(r_{ij}) \right]^2. \tag{7}$$

The explicit forms of these weight functions limit the range of the forces. At the inception of the DPD method, the classical form of the conservative weight function was chosen more for computational convenience rather than physical realism[7] and was given by

$$\omega^C(r_{ij}) = \omega^R(r_{ij}) \equiv \omega(r_{ij}) \tag{8}$$

and

$$\omega(r_{ij}) = \begin{cases} 1 - r_{ij}^* & for \ \ r_{ij}^* < 1 \\ 0 & for \ \ r_{ij}^* \geq 1 \end{cases}, \tag{9}$$

where, $r_{ij}^* = r_{ij}/R_C$, with $R_C$ the range of the force, but some time later numerical studies showed that an effective potential like the DPD conservative potential, was well defined[13].

The equations (6) and (7) mean that each particle feels Brownian noise and a friction force that acts in such a way that mass and momentum are conserved, therefore hydrodynamics is fulfilled.

Two additional conservative forces are incorporated, one to model polymers given by

$$\boldsymbol{F}_{ij}^{spring} = -K(r_{ij} - r_{eq})\hat{\boldsymbol{e}}_{ij}, \tag{10}$$

the constants $K = 100.0$ and $r_{eq} = 0.7$ have been chosen in such way that connected beads do not behave as independent particles, nor the computational efficiency of the method is affected. Additionally, the $r_{eq}$ value is chosen as the first maximum of the polymer radial distribution function if the spring constant is set to zero[14].

Another additional conservative force is incorporated to model, the electrostatic interactions $F_E$ that is explained in detail in the next section. These two additional forces are summed to the total force in equation (1).

The classical GCMC is well known[15] and consist in the exchange particles with a virtual bulk, adding or removing those particles of the simulation box with equal probability. The positions and velocities are chosen randomly; finally the resultant configuration is subjected to the Metropolis algorithm criterion[15] to decide whether this new configuration is accepted or not.

The difference of the method described above with the hybridized (DPD-GCMC) consists of the use of the DPD forces to calculate the positions and velocities of the particles instead of choosing them randomly; this hybrid scheme is explained in full detail in reference 12.

## II. EWALD SUMS AND CHARGE DISTRIBUTION

The Ewald sums technique is a method for computing long-range contributions to the potential energy in a system with periodic boundary conditions, like the electrostatic potential in an *N*-particles system. Hence, it is a technique for efficiently summing up interactions between an ion and all its periodic images. This method is the most employed route to calculate electrostatic interactions in molecular simulation[15]. It was originally developed in the study of ionic crystals[16].

Let us consider a simple cubic lattice of $L$ spacing. In each cell there are $N$ particles every one carrying a charge $q_i$. In the cell centered at $\boldsymbol{n}$ the particles with charge $q_i$ are at $\boldsymbol{r_i} + \boldsymbol{n}$ (periodic boundary condition). The total electrostatic contribution to the potential energy of the $N$ particles in the basic simulation cell (a single cell) can be written in SI units as

$$U_E(\boldsymbol{r}^N) = \frac{1}{4\pi\varepsilon_0\varepsilon_r}\sum_{\boldsymbol{n}}{}' \left(\sum_{i=1}^{N-1}\sum_{j>i}^{N}\frac{q_i q_j}{|\boldsymbol{r}_{ij} + \boldsymbol{n}L|}\right), \tag{11}$$

where $\varepsilon_0$ and $\varepsilon_r$ are dielectric constants of vacuum and solvent at room temperature, respectively, $\boldsymbol{n} = (n_xL, n_yL, n_zL)$, $n_x, n_y, n_z$, are integer numbers and the prime on the lattice sum indicates that if $\boldsymbol{n} = \boldsymbol{0}$ the $i = j$-terms are to be omitted, this sum over $\boldsymbol{n}$ takes into account the periodic images. It can be shown[17] that equation (11) can be written as Ewald sums:

$$U_E(\boldsymbol{r}^N) = \frac{1}{4\pi\varepsilon_0\varepsilon_r}\left[\sum_{i=1}^{N-1}\sum_{j>i}\left(q_i q_j \frac{erfc(\alpha r_{ij})}{r_{ij}} + \frac{2\pi}{V}\sum_{\boldsymbol{k}\neq 0}^{\infty} q_i q_j \frac{exp\left(-\frac{k^2}{4\alpha^2}\right)}{k^2} cos(\boldsymbol{k}\cdot\boldsymbol{r}_{ij})\right) - \frac{\alpha}{\sqrt{\pi}}\sum_{i=1}^{N} q_i^2\right], \tag{12}$$

This potential contains a real space sum (first term in equation (12)), plus a reciprocal space sum (second term), minus a self – interaction term (last term). In real space sum the term $erfc(x)$ is the complementary error function, $\alpha$ is the parameter that modulates the contribution of the sum in real space. As we can see from equation (9), the soft core interaction of the DPD beads allows the collapse of point charges and the subsequent formation of artificial ionic pairs. To avoid this inconvenience we approximate the charge as a distribution with spherical symmetry and an exponential Slater-type radial decay of length $\lambda$, following reference 18:

$$\rho(r) = \frac{q}{\pi\lambda^3}e^{-\frac{2r}{\lambda}}, \tag{13}$$

with

$$q = 4\pi\int_0^{\infty}\rho(r)r^2 dr. \tag{14}$$

Using the classical procedure to find the electrostatic potential between two charged distributions (Gauss's law) does not lead to solutions that can be calculated analytically in general. However some authors have found approximate expressions that give good accuracy at all relevant distances[18] for charge distributions given by (13):

$$U_E(r_{ij}, q_i, q_j) = \frac{q_i q_j}{r_{ij}} \left[ 1 - \left( \frac{\lambda_i \lambda_j + (\lambda_i + \lambda_j)^2}{(\lambda_i + \lambda_j)^3} r_{ij} + 1 \right) \times exp\left( -2 \frac{\lambda_i \lambda_j + (\lambda_i + \lambda_j)^2}{(\lambda_i + \lambda_j)^3} r_{ij} \right) \right], \quad (15)$$

we also consider all the charge distributions with the same decay length $\lambda = \lambda_i = \lambda_j$, hence the potential is given by:

$$U_E(r_{ij}) = \frac{q_i q_j}{r_{ij}} \left[ 1 - \left( \frac{5}{8} \frac{1}{\lambda} r_{ij} + 1 \right) \times exp\left( -2 \frac{5}{8} \frac{1}{\lambda} r_{ij} \right) \right], \quad (16)$$

defining

$$\beta \equiv \frac{5}{8} \frac{1}{\lambda} \quad (17)$$

thus

$$U_E(r_{ij}) = \frac{q_i q_j}{r_{ij}} \left[ 1 - (\beta r_{ij} + 1)\, e^{(-2\beta r_{ij})} \right]. \quad (18)$$

To compare with others, we use DPD reduced units, where $R_c^* = (k_B T)^* = m^* = 1$. Thus, the reduced electrostatic potential between two charge distributions with valence $Z_i$ and $Z_j$ respectively, separated by a distance $r^* = r_{ij}/R_c$ is given by

$$\frac{4\pi U_E^*(r^*)}{\Gamma} = \frac{Z_i Z_j}{r^*} \left[ 1 - (1 + \beta^* r^*) e^{-2\beta^* r^*} \right] \quad (19)$$

where $\Gamma = e^2/(k_B T \varepsilon_0 \varepsilon_r R_c)$ and $\beta^* = R_c \beta$. The value $R_c = 6.46$ Å is used, following Gonzalez-Melchor et al.[19]. The Coulombic term ($\sim \frac{1}{r^*}$) is evaluated then via the Ewald sums. Figure 1 shows the interaction between two charge distributions calculated with this method and the conservative DPD potential.

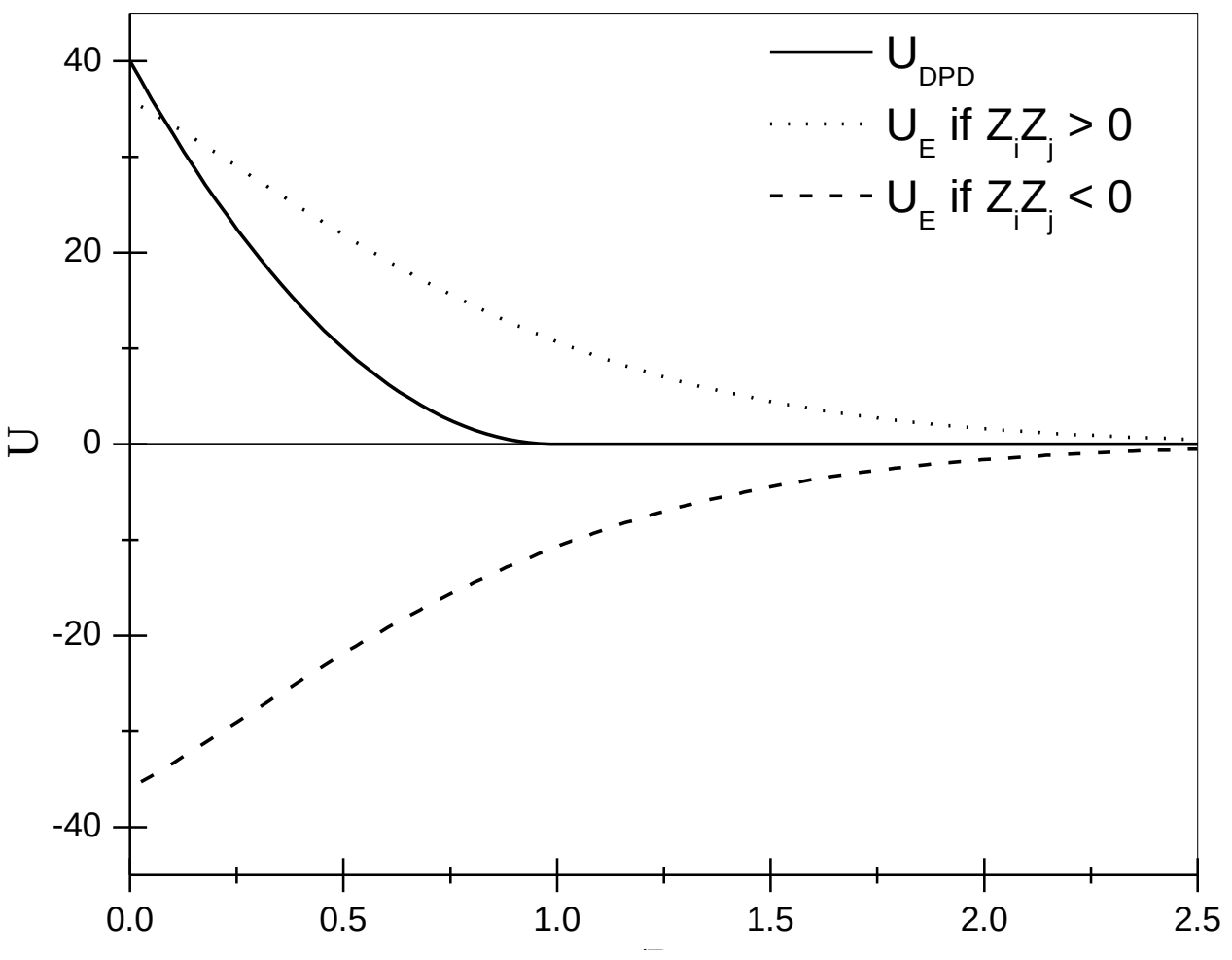

**Figure 1 Electrostatic potential using Ewald sums with Slater type charge distributions with $Z_iZ_j > 0$ (dotted line), $Z_iZ_j < 0$ (dashed line) both cases with $|Z_i| = |Z_j| = 5$ and the DPD conservative potential with $a_{ij} = 78.3$ (solid line).**

Since the electrostatic potential is conservative, the Coulomb force is given by

$$\boldsymbol{F}_E^*(r^*) = -\nabla U_E^*(r^*) \tag{20}$$

hence, from the potential of equation (19) we get

$$\frac{4\pi F_E^*(r^*)}{\Gamma} = \frac{Z_i Z_j}{r^{*2}}\left\{1 - [1 + 2\beta^* r^*(1 + \beta^* r^*)]e^{-2\beta^* r^*}\right\}. \tag{21}$$

The force in equation (21) is added to the DPD conservative force to give us the total conservative force which will determine the thermodynamic behavior of the system[20].

## III. VALIDATION OF THE ELECTROSTATIC MCGC – DPD

In order to validate our method we carried out simulations of systems taken from the literature[19,21] and calculated the radial distribution function (RDF) that determines the structure of the elements in the fluid. In both references they used canonical ensembles so the number of particles is constant. They simulated a fluid with 3000 DPD beads in a simulation box of volume $V^* = 10 \times 10 \times 10$; which leads a total number density of $\rho^* = 3$. To make a direct comparison with their results we used a chemical potential $\mu^* = 13.1$ that has been determined by DPD simulations of a fluid of one monomeric species with a repulsion parameter of $a_{ii} = 25.0$[12] hence the mean of the number density is $\langle \rho^* \rangle \sim 3$ in a simulation box of volume $V^* = 10 \times 10 \times 10$ so the number of particles is around 3000. 98 particles represent ions with net charge $e$ and the counterions are represented by 98 particles with net charge $-e$, the rest of the particles represent the solvent. We used the same set of parameters as references 19 and 21; the conservative interaction parameter was $a_{ij} = 25.0$ for all pairs of particles, the parameters for the dissipative and random forces intensities were $\gamma = 4.5$ and $\sigma = 3.0$, respectively; using the reduced thermal energy of $k_B T^* = 1$, a reduced time step of $\Delta t^* = 0.02$ was used. The parameters for Ewald sums were $\alpha = 0.15$ Å$^{-1}$ and the maximum vector $\mathbf{k}^{\max} = (5,5,5)$. For the charge distribution the value of $\beta^* = 0.929$ was used as in reference 19. For each simulation, $10^5$ MC configurations were generated and for each MC configuration 10 DPD moves were applied where the movements were integrated numerically under the velocity Verlet algorithm for DPD particles[22]. It is important to stress that only solvent particles were exchanged during the MCGC algorithm because the electroneutrality must be kept in every step of the simulation. In Figure 2 we show the resultant RDF's for the system explained before and the comparison with previous studies[19,21].

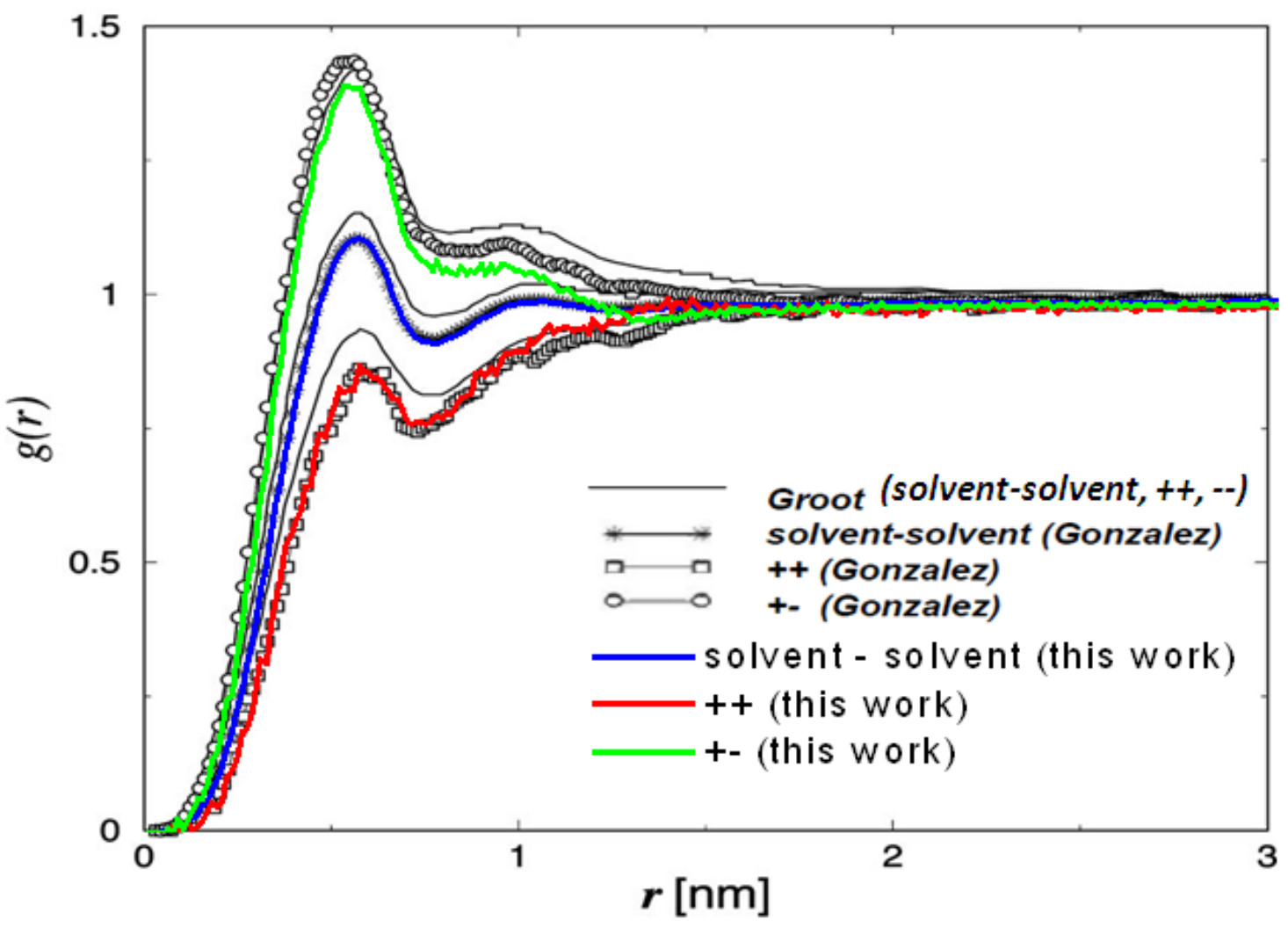

**Figure 2 Comparison between pair-correlation functions obtained by Groot[21], Gonzalez-Melchor et al.[19], and this work for a bulk electrolyte.**

The results in Figure 2 show general good agreement between all the methods, particularly our results have an almost perfect agreement between the solvent's RDF of Gonzalez-Melchor et al.'s, which were obtained in a Canonical ensemble[19] and this work, in spite of the difference of ensemble. Thus, the fluctuations in the number of solvent particles imposed by the GC ensemble do not affect the general structure of the solvent. However, we can see that the RDF's for the charged particles are slightly different from those of reference 19; this change is related also to the fluctuations in the number of solvent particles.

An additional comparison is given by the simulation of a polyelectrolyte – surfactant solution. We compare again our results with those reported in references 19 and 21, for a similar system. The simulations in these references used a fix number of 10125 particles in a cubic simulation box of volume $V^* = 15 \times 15 \times 15$. In our case, we set the chemical potential to $\mu^* = 13.1$ so that one gets approximately the same number density $\langle \rho^* \rangle \sim 3$. From this total number of particles, 50 beads represent 50 charged monomers of a polymer, each monomer carries a charge of *e/2*; surfactant molecules were modeled as 75 head – tail dimmers. The heads are represented by 75 charged beads, each bead with a net charge of –$e$, and tails are represented by 75 neutral beads. In order to preserve electroneutrality in the system 25 polymer counterions of charge –$e$ and 75 surfactant counterions of charge $e$ were added. The rest of the particles (around 9825) represent the water molecules. Polyelectrolyte and the surfactant molecules were connected by harmonic force with a spring constant of $K = 4.0$ and the bond equilibrium distance of zero. The repulsion interaction parameters were obtained using the expression[23]

$$a_{ij} = a_{ii} + \frac{\chi_{ij}}{0.306}, \tag{22}$$

where $\chi_{ij}$ is the Flory-Huggins interaction parameter[20], with $\rho^* = 3$ and $a_{ii} = 25$. The Flory-Huggins interaction parameter between tail beads ($c$) and water ($w$) is given by $\chi_{cw} = 18$; the head group ($h$) has the parameters $\chi_{ch} = 6$ and $\chi_{wh} = 0$; the polymer beads ($p$) have a solubility parameter $\chi_{pw} = 0.65$. This parameter is related to the noncharged polymer in a $\theta$ solvent[21], the other parameters are $\chi_{cp} = 6$ and $\chi_{ph} = 0$. The parameters for the dissipative and random force intensities were $\gamma = 4.5$ and $\sigma = 3.0$, respectively. Using the reduced thermal energy of $k_B T^* = 1$, we generated $2 \times 10^5$ MC configurations, and for each MC configuration 10 DPD steps were applied with a time step of $\Delta t^* = 0.04$ using the velocity-Verlet algorithm[22]. The parameters related to the Ewald sums and the charge distribution were the same as for the simulation of the bulk electrolyte.

Figure 3 shows the RDF's of a polyelectrolyte-surfactant mixture in aqueous solution. We used the same set of parameters as those of references 19 and 21 to make direct comparisons. The agreement is good for polymer-polymer (pp) and polymer-head (ph) interactions. In particular, the RDF obtained with our method for head-head (hh) particles is more similar to Groot´s distribution function at long distances than Gonzalez-Melchor et al.'s[19]. The explanation for this difference given by these last authors is that it might be due to the absence of polarization effects in the electrostatic interaction that they used. However we used the same Ewald sums method as reference 19 and reproduce the same RDF as Groot's for large distances (see inset in Figure 3). Hence, we believe that the polarization does not play an important role in the RDF at large distances. We have the same values of RDF at large distance as Groot's because the DPD–MCGC method uses the Metropolis algorithm which keeps optimal configurations only. This is an advantage compared with simple dynamics.

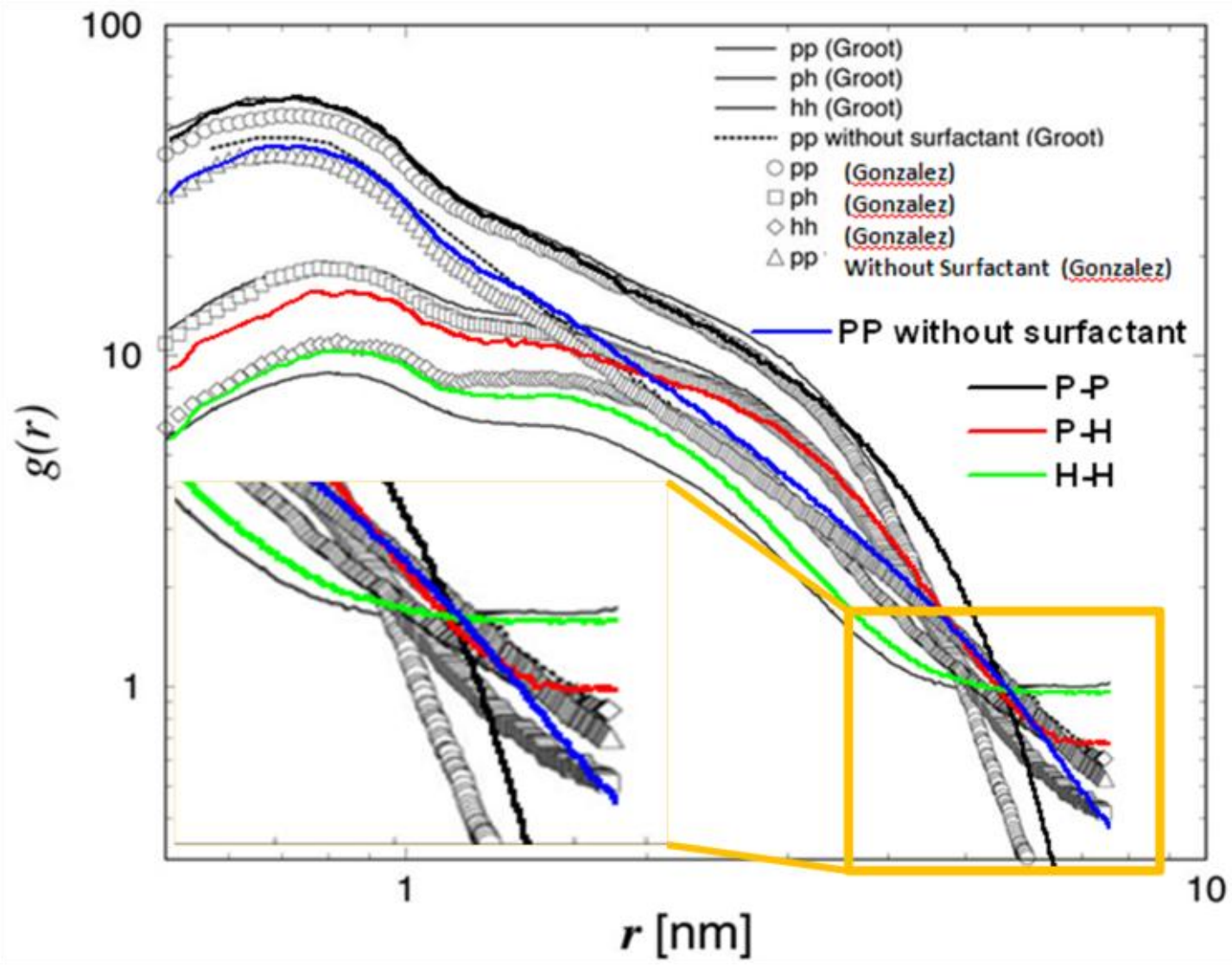

**Figure 3 Comparison between pair-correlation functions of an aqueous solution of polyelectrolyte-surfactant mixture obtained by Groot[21], Gonzalez-Melchor and co-workers[19] and us. The black curve is the intramolecular polymer-polymer correlation; polymer-head group is shown in red, head group-head group in green, and the blue one is the intramolecular polymer-polymer correlation with surfactant replaced by salt. The inset shows the comparison of the h-h curve predicted by us with those of references 19 and 21. See text for details.**

## IV. INFLUENCE OF THE COULOMB INTERACTION STRENGTH

To understand the effect of the electrostatic interaction in the conformational properties of the polyelectrolyte we performed some numerical simulations of salt-free polyelectrolyte solutions, taking into account the counterions and the solvent explicitly. It could be expected intuitively, that an increase in the number of the charges causes a stretching of a chain due to repulsion between the charged monomers. But, according to the Manning condensation theory[24] the counterions start to condense if

$$l_B \geq b, \tag{23}$$

with $l_B$ in SI units given by

$$l_B = \frac{e^2}{4\pi\varepsilon_0\varepsilon_r k_B T} \tag{24}$$

known as the Bjerrum length; $e$ is the unit charge and $\epsilon$ is the dielectric constant of the fluid, and $b$ is defined by

$$b = L/P \tag{25}$$

where $L$ is the contour length of the polyelectrolyte and $P$ the number of charged monomers[24].

In our simulations a normalized length was used as a measure of the interaction strength between charges:

$$\lambda = \frac{l_B}{R_C} k_B T. \quad (26)$$

$k_B T$ and $R_C$ are the DPD values for the thermal energy and the cutoff distance, respectively. The solvent quality is fixed to $\theta$ again by means of the repulsion parameter of the DPD forces, as done in the previous section. Polyelectrolyte conformation is then modulated by the quality of the solvent and the electrostatic forces among polyelectrolytes[25]. In these simulations, each monomer carries a charge $e$ and the counterions a charge $-e$. Hence, the charge of the total system is neutralized. We simulated two systems, one with a polyelectrolyte formed by 32 charged DPD monomers and another one formed by 8 charged DPD monomers, both at a concentration of $0.032/R_C^3$, and the volumes of the simulation boxes are $V* = 6.3 \times 6.3 \times 6.3$ and $V* = 10 \times 10 \times 10$, respectively. For both systems the repulsion parameters are $a_{ij} = 78.3$ for all particle pairs. The chemical potential was fixed at $\mu^* = 37.8$ so that the mean density of the fluid was $\langle \rho^* \rangle \sim 3$. The parameters for the dissipative and random forces intensities were $\gamma = 4.5$ and $\sigma = 3.0$, respectively. The spring constant for bonded monomers was of $K = 100.0$ and bond equilibrium distance of 0.7. Using the reduced thermal energy of $k_B T^* = 1$, we generated $10^6$ MC configurations and for each MC configuration 10 DPD steps were applied with a time step of $\Delta t^* = 0.02$. The parameters related to the Ewald sums and the charge distribution were the same as in the simulation of the bulk electrolyte and polyelectrolyte – surfactant solutions of Section III. The interaction strength, equation (26), covers an interaction range $0 \leq \lambda \leq 16.0$, that is, from a polymer without charges to an extensively charged polymer. The conformation that takes the polyelectrolyte for different strengths was determined with the squared radius of gyration, which is used to describe the polymeric conformation[26].

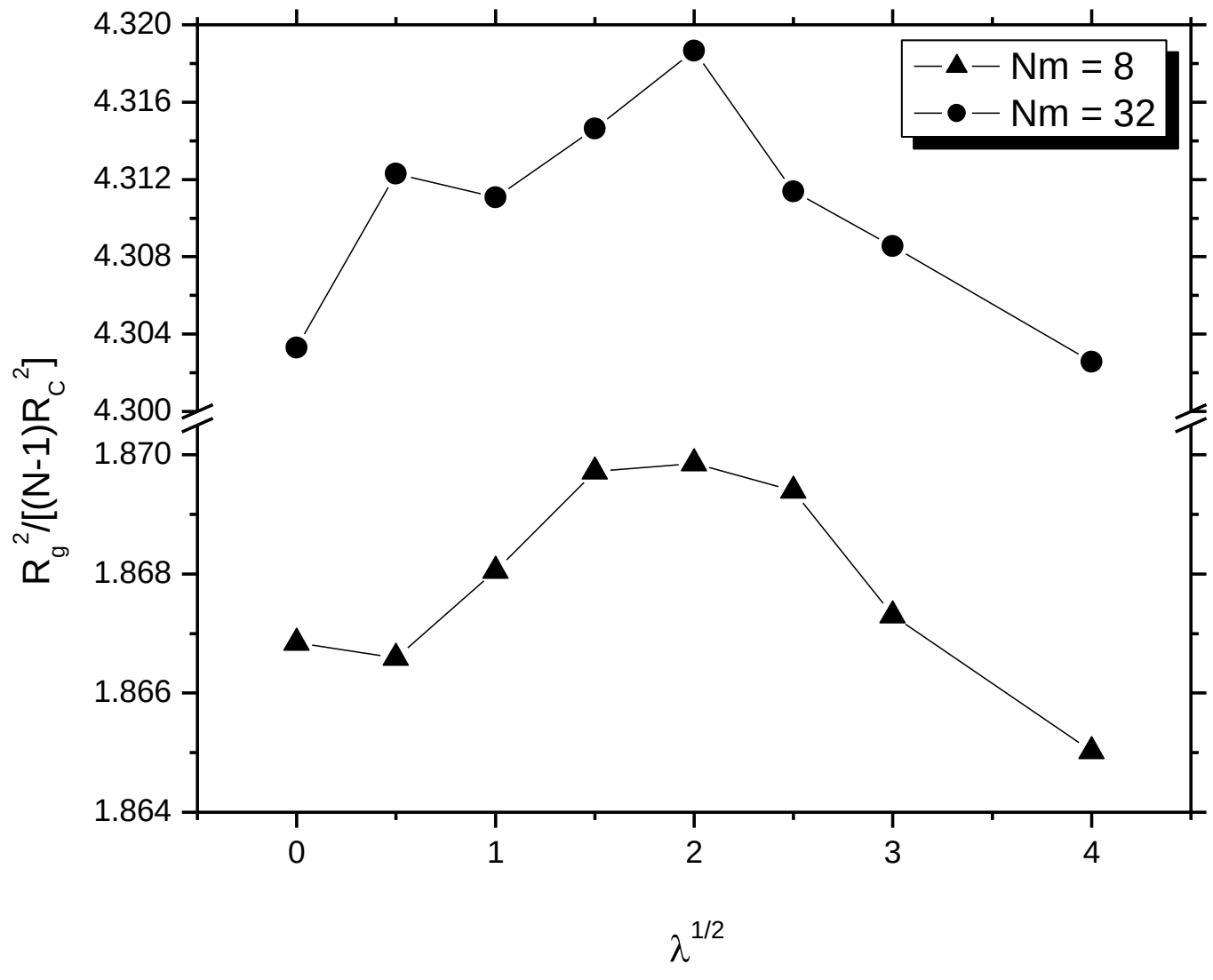

**Figure 4 Squared radius of gyration, divided by the number of bonds as a function of the square root of the interaction strength, for *N* = 8 (triangles) and *N* = 32 (diamonds). Error bars are smaller than the symbol size. Lines are guides to the eye.**

Figure 4 shows the behavior of the square radius of gyration of a polyelectrolyte as a function of the electrostatic interaction strength for $N = 8$ and $N = 32$. For small values of $\sqrt{\lambda}$, with lambda given by equation (26), an expansion of the polyelectrolyte due to the electrostatic repulsion between monomers is obtained, until a maximum is reached at the same value of the interaction strength for both lengths of the polyelectrolyte. This is the critical charge spacing needed to induce partial screening of the charges of the chain. Further increase of the interaction strength leads to chain contraction, due to the raise in the number of counterions condensed on the polyelectrolyte. These condensed counterions induce a reduction in the electrostatic repulsion between monomer beads since the counterions form ionic pairs with the monomers, and the result is the collapse of the polyelectrolyte. We do find an expansion of the polyelectrolyte due to the interaction strength and a collapse due to the excess in the strength of the electrostatic interactions.

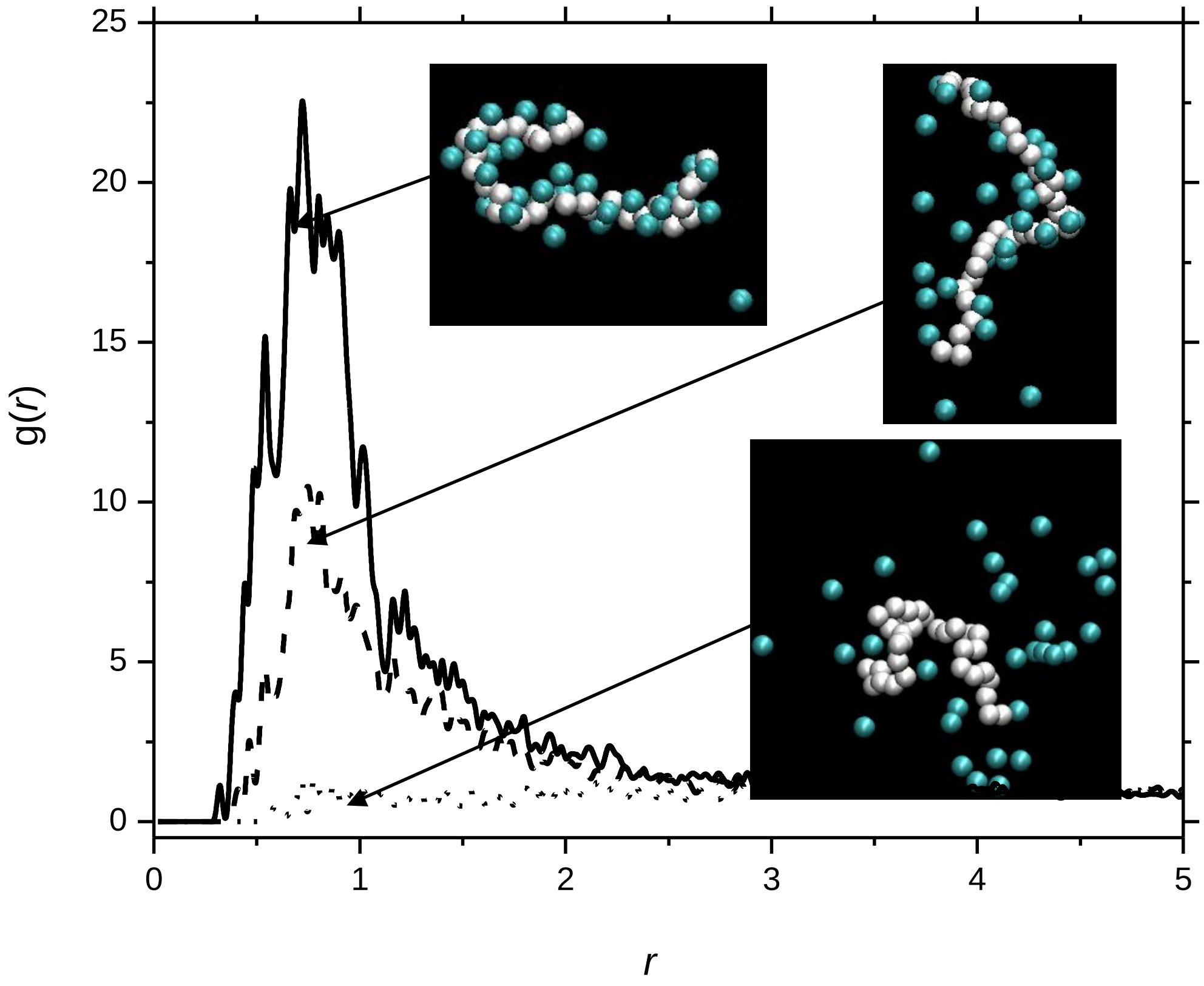

**Figure 5 Radial distribution functions between the charged monomers of a single chain formed by 32 beads and their counterions. The functions are complemented by snapshots[27] of the polyelectrolyte (white beads) and the counterions (blue beads), where the solvent beads are not shown for clarity. For various values of interaction strength, $\sqrt{\lambda}$ = *0* (dotted line), $\sqrt{\lambda}$ = *2* (dashed line) and $\sqrt{\lambda}$ = *4* (solid line).**

To observe the transition of the polyelectrolyte conformation shown in Figure 4 we present the RDF's of the monomers with respect to their counterions for three distinctive values of interaction strengths $\sqrt{\lambda} = \{0, 2, 4\}$ in Figure 5. With no electrostatic interactions ($\sqrt{\lambda} = 0$), the conformation it takes arises from the DPD interactions only, and represents a noncharged polymer in $\theta$ solvent. The counter ions are thus uniformly distributed around the polymer as a consequence of the DPD forces only; the RDF and the snapshot[27] confirm this argument. For $\sqrt{\lambda} = 2$ (the critical point in Figure 4), the counterions are condensed on the polyelectrolyte backbone, extending the chain. Finally, for $\sqrt{\lambda} = 4$ the polyelectrolyte assumes a less extended configuration, similar to the one taken when there are no electrostatic interactions ($\sqrt{\lambda} = 0$). However, the distribution of the counterions around the polyelectrolyte is very different, as there are virtually none free in the solution and most have condensed on the polymer sites. Undoubtedly, at this high strength the collapse is due to the screening of the monomers charge, which makes the Coulomb interaction secondary to the fact that the polymer is in $\theta$-solvent conditions. This is quantitatively seen in Figure 6.

A quantitative description of the counterion distribution around the polyelectrolyte is given by the net charge distribution around the polyelectrolyte. The net charge at a distance of one monomer of the polyelectrolyte is given by:

$$Q_T(r) = \sum_{i=1}^{N_s} Z_i \, g_i(r) \tag{27}$$

where $N_s$ is the number of species in the fluid, $Z_i$ the valence of the $i$- species and $g_i(r)$ the RDF between monomers and the molecules of the $i$-species.

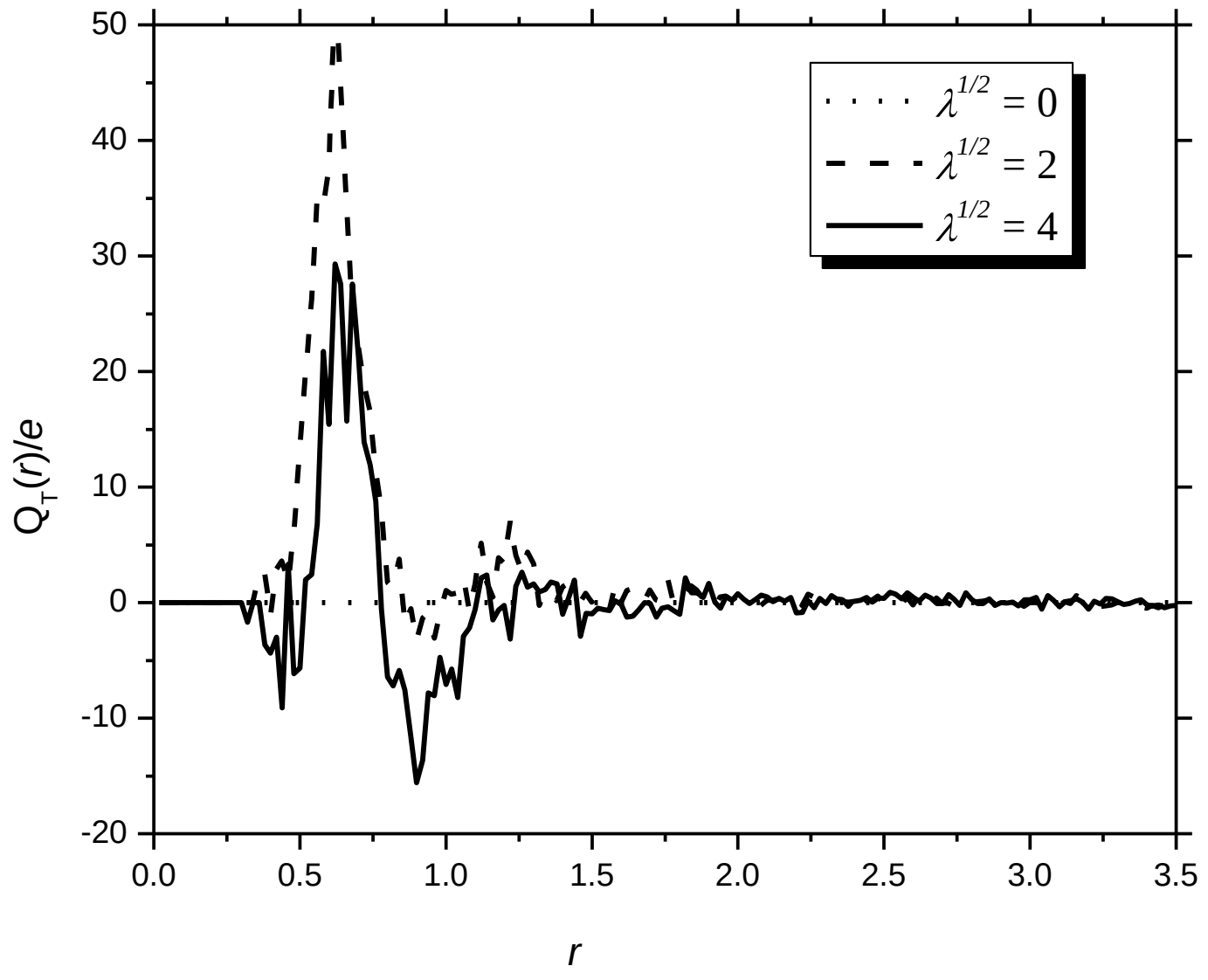

**Figure 6 Net charge distribution around the polyelectrolyte with $N$ = 32 for different interaction strengths, $\sqrt{\lambda}$ = *0* (dotted line), $\sqrt{\lambda}$ = *2* (dashed line) and $\sqrt{\lambda}$ = *4* (solid line).**

If the system has no interaction strength ($\sqrt{\lambda}$ = 0) there are no electrostatic charges, hence $Q_T(r)$ = 0 (dotted curve in Figure 6). Now, if we have an interaction strength of $\sqrt{\lambda}$ = 2, the monomers experience a stronger repulsion with each other than attraction with the counterions, resulting in an expansion of the polyelectrolyte. When the strength of the electrostatic interaction is increased to $\sqrt{\lambda}$ = 4, the net positive charge around monomers is reduced (solid curve in Figure 6), leading to a built-up of negative net charge for distances $r$~1, which in turn induces the polymer contraction.

## V. SALT–INDUCED COLLAPSE AND REEXPANSION OF POLYELECTROLYTES.

We performed a set of simulations of a model anionic polyelectrolyte, represented by a $N$-beads polyelectrolyte, with each bead carrying a charge of $-e$ and chain lengths equal to $N$ = *16, 32, 64, 96* to study the behavior of a polyelectrolyte in the presence of multivalent salt. In this case, monomers were bonded with a spring constant of $K$= 100.0 and bond equilibrium distance was of 0.7. The volumes of the simulation boxes were varied depending on the size of the polymer: $V^*$ = 390.24, 780.5, 1561, 2341.5 to keep the monomer concentration fixed at ${C_m}^* =$ 0.04 in each simulation. To preserve electroneutrality in the system, $N$ polymer counterions of charge $e$ were added; for each tetravalent ion four counterions were also added. The electrostatic effect is well observed with the repulsion interaction parameters $a_{ij}$ = 78.3, which represents a $\theta -$ solvent in the absence of charge. The chemical potential was $\mu^*$ = 37.8 so that the mean density $\langle\rho^*\rangle$ ~ 3. The parameters for the dissipative and random forces intensities were $\gamma =$ 4.5 and $\sigma =$ 3.0, respectively. Using the reduced thermal energy of $k_B T^* =$ 1, we generated $10^6$ MC configurations and for each MC configuration 10 DPD steps were applied with a time step of $\Delta t^* =$ 0.02. The parameters related to the Ewald sums and the charge distribution were the same as for the simulation of the bulk electrolyte and polyelectrolyte – surfactant solutions of section III.

We present the squared radius of gyration in Figure 7 for a linear, fully charged, $N$-bead polyelectrolyte in the presence of *(4:1)* salt, where 4 means salt valence. We use a tetravalent salt because it has been seen that for multivalent salt the effect of change in the conformation of the polyelectrolyte is more evident than for the monovalent one in atomistic simulations[28].In absence of salt, an extended conformation is observed as consequence of the electrostatic repulsion between monomers. Upon addition of salt, a decrease in ${R_g}^2$the ions screening of the Coulomb interaction, but a striking effect occurs once the salt concentration is increased beyond the concentration where the salt cations neutralize the bare polyelectrolyte charge and the chain starts to swell again.

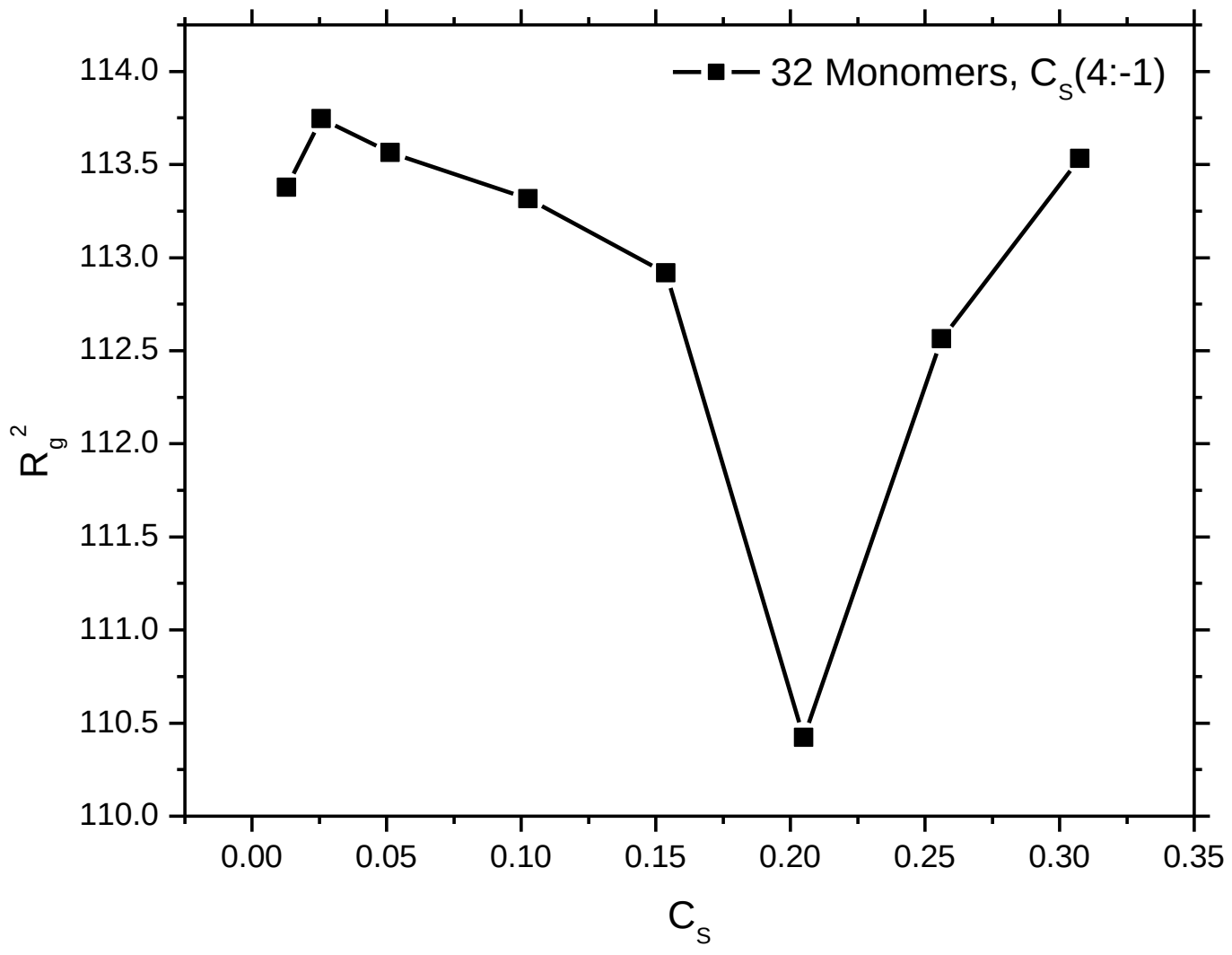

**Figure 7 Squared radius of gyration $R_g^2$ of a polyelectrolyte of length $N$ = 32 as a function of tetravalent salt concentration $Cs$. Error bars are smaller than the symbol size. Lines are guides to the eye.**

At low salt concentration ($C_s$ = 0.026) the polyelectrolyte is surrounded by the positive salt ions but the monomers are not compacted in certain position, as shown in Figure 8(a) (dotted line) where the definition of equation (27) has been used. When the salt concentration rises to $C_S$ = 0.205 (dashed line), the negatively charged monomers are compacted around $r$ ~ 0.75, and surrounding the polyelectrolyte there is a negatively charged cloud up to $r$ ~ 2.5, when the net charge distribution has a charge inversion, see dashed line in Figure 8(a). When the salt concentration rises even more than $C_S$ = 0.308 (solid line) the net charge inversion distribution starts at a shorter distance from the polyelectrolyte ($r$ ~ 2.0); this is due to the tetravalent salt ions that start to interact with their negative counterions that surround the polyelectrolyte. This pseudo ionic-pair formation is brought out as a consequence of the polyelectrolyte swelling. The RDF's between monomers and salt particles for $N$ = 32 and $C_S$ = {0.026, 0.205, 0.308}, depicted in Figure 8(b), provide additional information about the interaction between the multivalent salt and the polyelectrolyte. The RDF at $C_S$ = 0.026 tends to zero at large distance because there are few salt ions and most of them surround the polyelectrolyte at the maximum of the RDF, that is at $r$ ~ 0.86. However this condensation it is not enough to collapse the polyelectrolyte. At $C_S$ = 0.205 there are enough salt ions distributed around the polyelectrolyte so that three coordination shells (see dashed line Figure 8(b)) are formed, hence the polyelectrolyte takes a collapsed configuration. If the salt concentration rise to $C_S$ = 0.308 we can infer, based on the solid line in Figure 8(b), that the excessive amount of salt starts to interact more with the counterions than with the polyelectrolyte, whose charge is totally screened out and the third coordination shell that appears at lower salt concentrations vanishes. We conclude therefore that this latter, more fluidic, configuration of salt ions around the polyelectrolyte is responsible for the reexpansion of the polyelectrolyte.

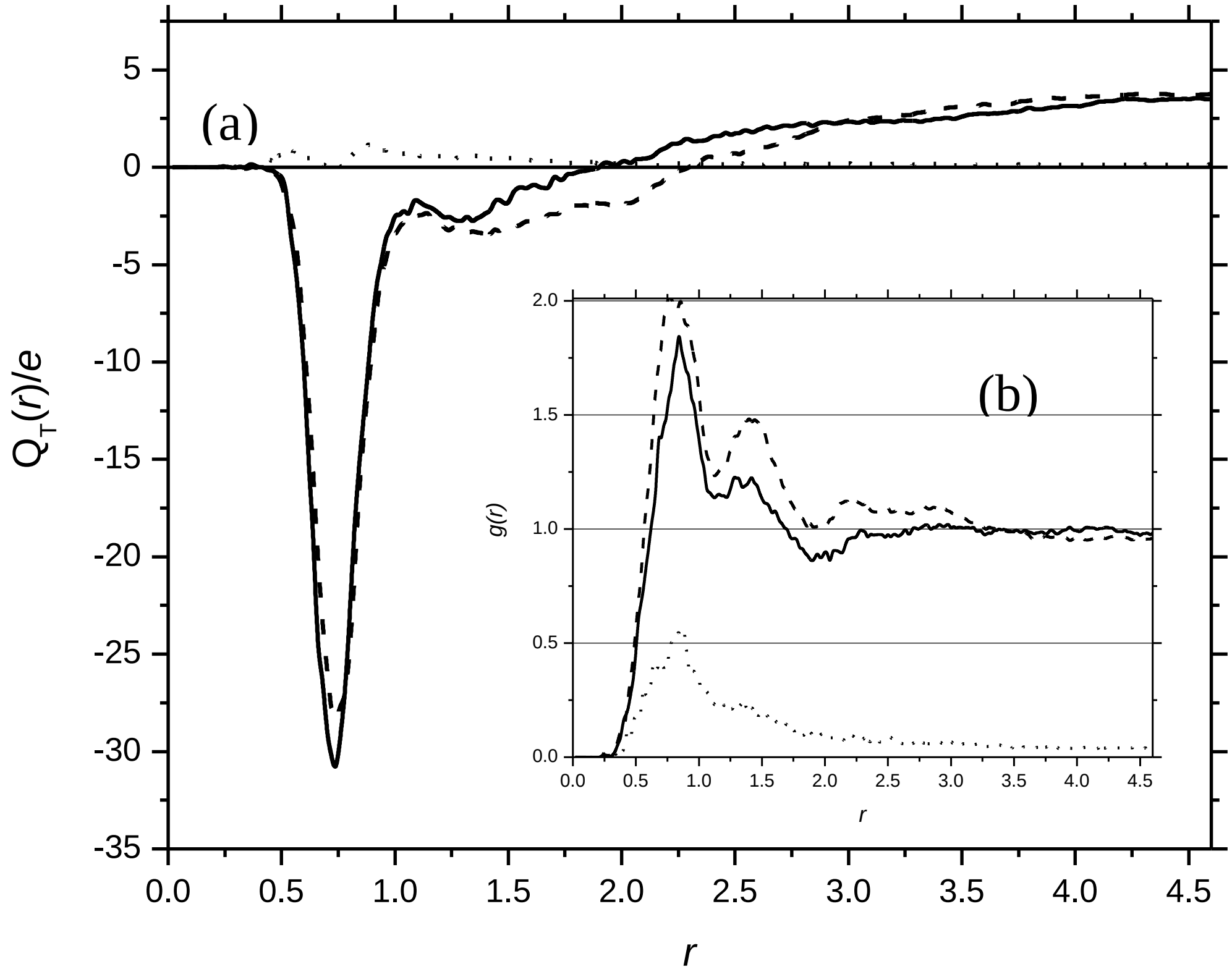

**Figure 8 (a) Net charge distribution around the polyelectrolyte with *N* = 32, for different salt concentrations. (b) The RDF's between polyelectrolyte and tetravalent salt with *N* = 32, at different salt concentrations. For both figures the curves represent $C_S$ = 0.026 (dots), $C_S$ = 0.205 (dash) y $C_S$ = 0.308 (solid).**

Figure 9 shows the same calculation as in Figure 7, but for a larger polyelectrolyte *(N = 64)*. In addition, we present $R_g^2$for a neutral polymer (circles curve) with the same simulation parameters. We can observe that the polyelectrolyte collapses due to the salt ions and the explicit solvent particles, which is not directly attributable to the electrostatic interactions between the salt and the charged monomers, even if collapse and reexpansion of the non-charged polymer is smaller than for the charged one. The collapse occurs at larger concentration than for the *N*=32 polymer because there are more charged sites on the polymer structure whose screening requires more salt ions. The collapse and reexpansion mechanisms are therefore similar to those in Figure 7.

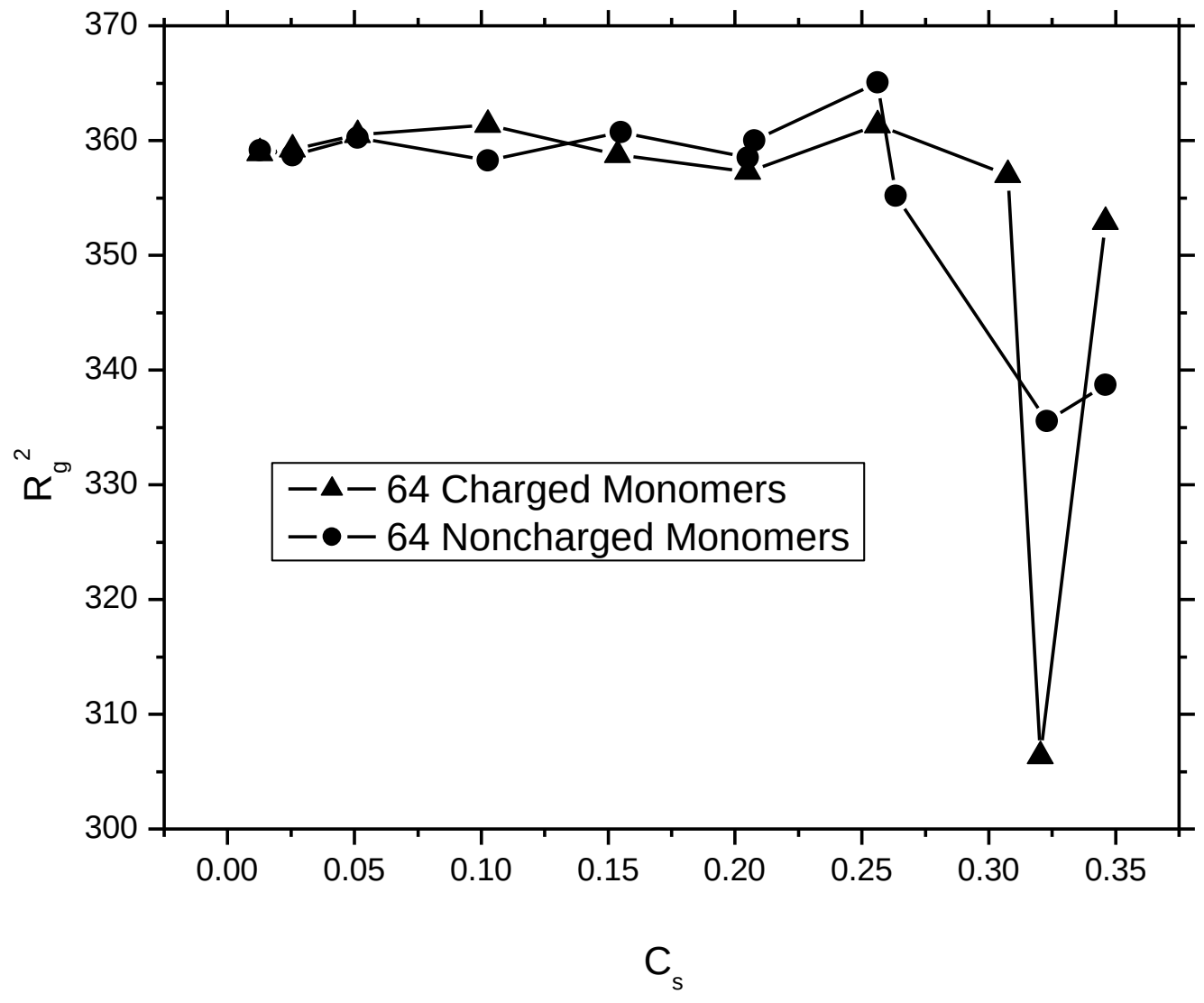

**Figure 9 Squared radius of gyration $R_g^2$ of a polyelectrolyte (triangles) and a neutral polymer (circles) with degree of polymerization $N$ = 64 as a function of tetravalent salt concentration Cs. Error bars are smaller than the symbol size. Lines are guides to the eye.**

We investigated the variation of the squared $R_g$ with degree of polymerization $N$ to quantify the swelling of the polyelectrolyte by tetravalent salt, assuming that the polyelectrolyte conformational follows a scaling behavior like the noncharged polymer[29], $R_g \sim N^{\nu}$.However, the situation is far more complicated for polyelectrolyte solutions where at least three components, e.g., polymer, counterion, and solvent, play an important role. It has been shown both by simulations and experimentally that the conformational properties depend not only on the charge on the polymer and the concentration of added salt but also on the chemical nature of the salt ions[28,30]. Nevertheless we have found that a scaling exponent ν can still be extracted from these systems, as depicted in Figure 10, where the critical exponent ν as a function of the salt concentration is shown.

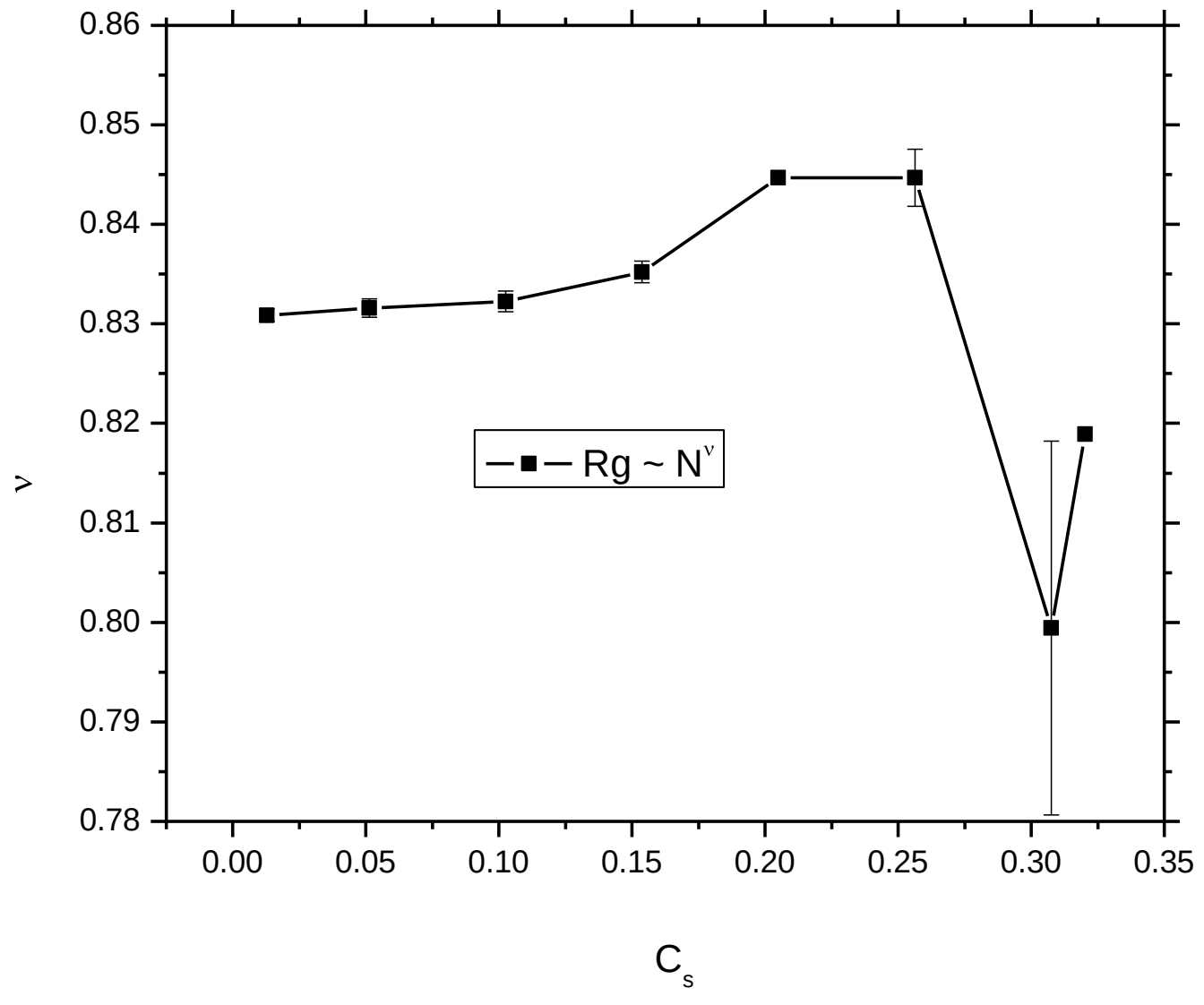

**Figure 10 Scaling exponent ν of the radius of gyration for a polyelectrolyte in a tetravalent salt solution. Statistical uncertainties are indicated for each data point. Lines are guides to the eye.**

The values of the critical exponent $\nu$ have been obtained through logarithmic fits to the squared radius of gyration data for all the polyelectrolyte sizes modeled, $N$ = {*16*, *32*, *64* y *96*}, at each salt concentration. The values for ν are found to be between 0.78 and 0.85, which means that the polyelectrolyte is fairly expanded. These results are in agreement with de Gennes[29] analysis that showed that $\nu \rightarrow 1$ for polyelectrolytes solutions due to the repulsion between monomers. Another important observation is that we used a mesoscopic method and the scale of the critical exponent is different from that obtained by others, for example in the atomistic simulations of Hsiao and Luijten[28] where the change in the critical exponent in function of the salt concentration is greater than in our simulations.

It is well known that the quality of the solvent is important in determining the structural conformation of polymers in solution[26]. With this in mind we carried out some simulations for a polyelectrolyte with $N$ = 32 at different salt concentrations, but modifying the quality of the solvent by means of the values of the DPD conservative force repulsion parameter (see equation (3)) to see its effect on the conformation of the polyelectrolyte. We define good solvent conditions in terms of the repulsion parameter as $a_{ij} < a_{ii}$, bad solvent for $a_{ij} > a_{ii}$ and $\theta$ solvent as already defined ($a_{ij} = a_{ii}$). The rest of the input parameters were the same as those for the set of simulations depicted in Figure 7.

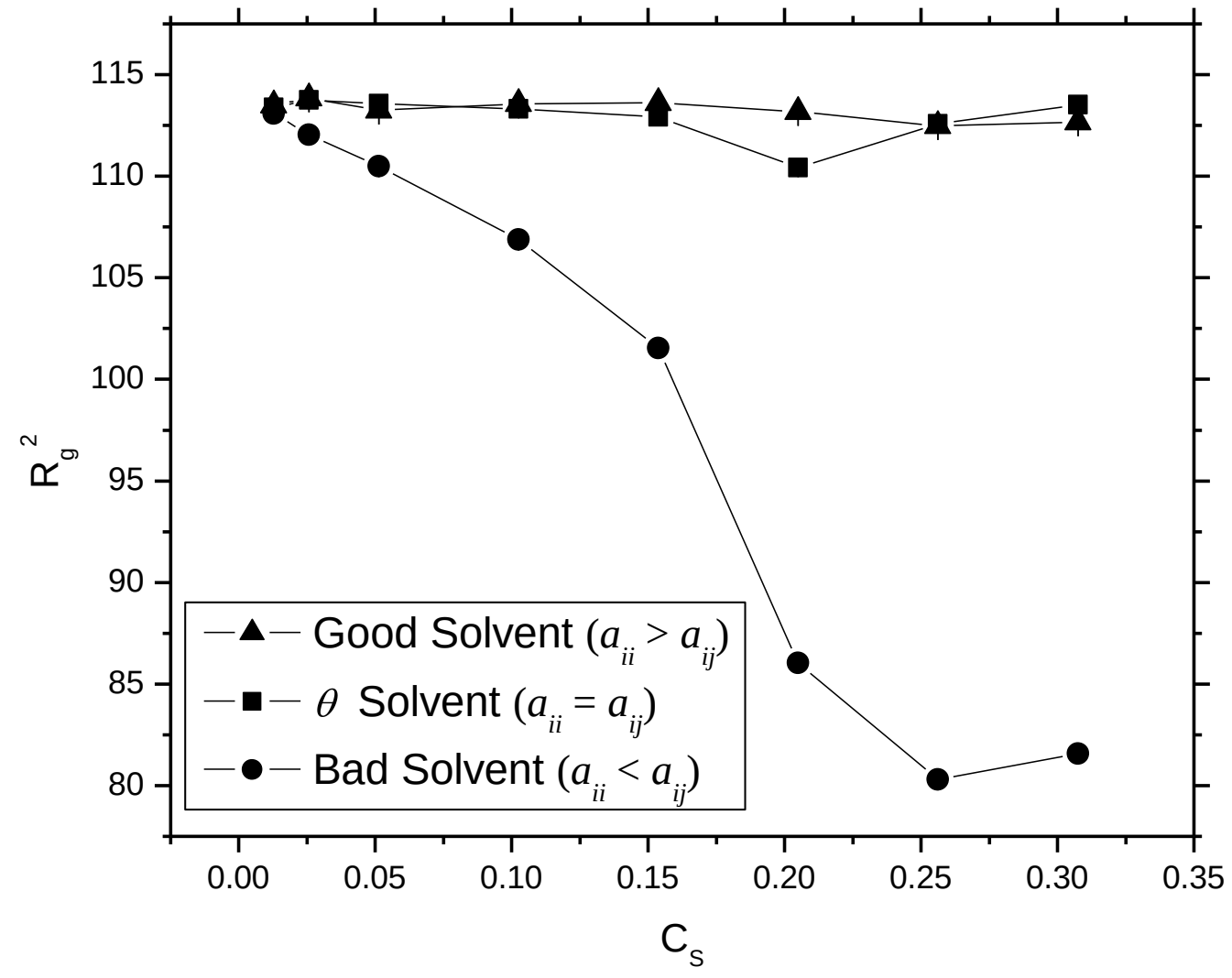

**Figure 11 Squared radius of gyration $R_g^2$ of a polyelectrolyte of length $N$ = 32 as a function of tetravalent salt concentration *Cs*, for different solvent quality. The results for $\theta$ solvent are the same as the Figure 7 (square symbols). Error bars are smaller than the symbol size. Lines are guides to the eye.**

Figure 11 shows the results for the squared radius of gyration of the polyelectrolyte, in the presence of salt, obtained for different solvent qualities. At low salt concentration the polyelectrolyte takes a more compacted conformation for bad solvent, as expected, than for good- or $\theta$-solvents. However, what might appear somewhat unexpected is that upon addition of salt, the change in the polyelectrolyte conformation for bad solvent is much greater than in the other cases. One must keep in mind though, that the range of the Coulomb interaction is decreased with increasing $C_S$, leaving the DPD repulsions as the leading interactions. Yet a reexpansion can still be observed, which we believe is of electrostatic origin (see discussion of Figure 7) hence the need for the explicit inclusion of the electrostatic interactions into the model.

## VI. SUMMARY AND CONCLUSIONS

Mesoscopic simulations using Monte Carlo hybridized with dissipative particles dynamics in the Grand Canonical ensemble were presented for the study of conformational properties of polyelectrolytes in solution, with and without polyvalent salt ions, with the inclusion of electrostatic interactions via Ewald sums. Good agreement with systems taken from the literature was obtained. We have also shown the collapse of the chain and the following reexpansion after some critical salt concentration, in qualitative agreement with results reported for good solvent using Lennard – Jones microscopic simulations[28]. Due to the mesoscopic character of our simulations the results for the critical exponent indicate that with our method the polyelectrolyte does not collapse as much as a neutral polymer, but one must keep in mind that our scale is greater than that of the atomistic simulations. Additionally, the explicit introduction of the solvent particles leads to solvent – induced attraction which is present even between interior sites of the collapsed polyelectrolyte, where no solvent molecules are present. A similar scenario occurs for the salt-free polyelectrolyte, where a counterion condensation is exhibited but not as strong as that reported by other authors using different methods and solvent qualities[31]. As far as we know these are the first mesoscopic simulations that calculate the radius of gyration as function of the electrostatic interaction strength or as a function of salt concentration. We have also showed the importance of the solvent quality in terms of the DPD repulsion parameter, in polyelectrolyte solutions in the presence of tetravalent salt.

One advantage of implementing an electrostatic version of DPD in the Grand Canonical ensemble is that it can be applied to systems confined by walls, where the chemical potential must be fixed to guarantee chemical equilibrium with a bulk, as is the case, for example, in experiments involving the surface force apparatus. Additionally, one could carry out studies on the influence of pH on the polyelectrolyte conformation. These results could be compared with experimental results like those obtained by Kirwan et al.[32], who reported the collapse of poly(vinylamine) molecules induced by changes in the solutions' pH.

**Acknowledgements**

This work was supported by the Centro de Investigación en Polímeros (CIP, COMEX Group). Some of the calculations were performed at the Centro Nacional de Supercómputo – IPICYT. The authors thank C. Narambuena for fruitful discussions.

**REFERENCES.**

1. A.V. Dobrynin, Current Opinion in Colloid & Interface Science **13**, 376 (2008).
2. A. Yethiraj, J. Phys. Chem. B **113**, 15390 (2009).
3. K. Huang, *Statistical Mechanics* (John Wiley & Sons, New York, 1987).
4. A. Gama Goicochea, E. Nahmad-Achar, and E. Pérez, Langmuir **25**, 3529 (2009).
5. F. Goujon, P. Malfreyt, and D. J. Tildesley, ChemPhysChem **5**, 457 (2004).
6. A. K. Gunstensen and D. H. Rothman, J. Geophys. Res. **98**, 6431 (1993).
7. P. J. Hoogerbrugge and J. M. V. A. Koelman, Europhys. Lett. **19**, 155 (1992).
8. P. Español and P. Warren, Europhys. Lett. **30**, 191 (1995).
9. I. Pagonabarraga and D. Frenkel, J. Chem. Phys. **115**, 5015 (2001).
10. P. Nikunen, M. Karttunen and I. Vattulainen, Comput. Phys. Commun. **153**, 407 (2003).
11. P. Español and M. Revenga, Phys. Rev. E. **67**, 026705 (2003).
12. A. Gama Goicochea, Langmuir ***23***, 11656 (2007).
13. B.M. Forrest and U.W. Suter, J. Chem. Phys. **102**, 18 (1995).
14. A. Gama Goicochea, M. Romero-Bastida and R. López-Rendón, Mol. Phys. **105**, 17 (2007).

15. D. Frenkel and B. Smit, *Understanding Molecular Simulations, From Algorithms to Applications* (Academic, New York, 1996).
16. P. P. Ewald, Ann. Phys. **64**, 253 (1921).
17. S.W. de Leeuw, J.W. Perram and E.R Smith, Proc. R. Soc. London A **373**, 27 (1980).
18. M. Carrillo-Tripp, H. Saint-Martin, and I. Ortega-Blake, J. Chem. Phys. **118**, 7062 (2003).
19. M. González-Melchor, E. Mayoral, M. E. Velázquez and J. Alejandre, J. Chem. Phys. **125**, 224107 (2006).
20. R. D. Groot and P. B. Warren, J. Chem. Phys. **107**, 4423 (1997).
21. R. D. Groot, J. Chem. Phys. **118**, 11265 (2003).
22. I.Vattulainen, M. Karttunen, G. Besold and J. M. Polson J. Chem. Phys. **116**, 3967 (2002).
23. A table with the explicit values of the interaction parameters can be found in reference 19.
24. G. S. Manning, J. Chem. Phys. **51**, 924 (1969).
25. D. H. Napper, *Polymeric Stabilization of Colloidal Dispersions* (Academic Press, London, 1983).
26. Flory P.J. *Principles of Polymer Chemistry* (Cornell University, New York, 1953).
27. W. Humphrey, A. Dalke and K. Schulten, J. Molec. Graphics **14.1**, 33 (1996).
28. P.Y. Hsiao, E. Luijten, Phys. Rev. Lett. **97,** 148301 (2006).
29. P.G. de Gennes, *Scaling Concepts in Polymer Physics*, (Cornell University Press, Ithaca, NY, 1979).
30. N. Volk, D. Vollmer, M. Schmidt, W. Oppermann, K. Huber, Adv. Polym. Sci. **29**, 166 (2004).
31. R.G. Winkler, M. Gold and P. Reineker, Phys. Rev. Lett. **80**, 3731 (1998).
32. L. J. Kirwan, G. Papastavrou, M. Borkovec, Nano Lett. **4**, 149 (2004).